\documentclass[12pt,1p  ]{elsarticle}

\usepackage{amssymb}
\usepackage{amsmath}
\usepackage{graphics}
\usepackage{latexsym}
\usepackage{array}
\usepackage[activeacute,english]{babel}
\usepackage[latin1]{inputenc}

\usepackage{color}
\usepackage{lineno}
\usepackage{latexsym}
\usepackage[colorlinks=true]{hyperref}

\newcommand{\bn}{\begin{eqnarray}}
\newcommand{\en}{\end{eqnarray}}
\newcommand{\bq}{\begin{equation}}
\newcommand{\eq}{\end{equation}}

\newcommand{\bc}{\begin{center}}
\newcommand{\ec}{\end{center}}

\journal{Annals of Physics}

\begin{document}

\begin{frontmatter}

\title{Exact solutions of $(n + 1)$-dimensional
Yang-Mills equations in curved space-time}

\author[rvt]{J. A. S\'anchez-Monroy\corref{cor1}}
\ead{antosan@gmail.com}
\author[rvt,focal]{C. J. Quimbay\corref{cor2}}
\ead{cjquimbayh@unal.edu.co}

\address[rvt]{Grupo de Campos y Part\'\i culas, Universidad Nacional de Colombia, Bogot\'a D.C., Colombia}
\address[focal]{Associate researcher of Centro Internacional de F\'{\i}sica, Bogotá D.C., Colombia}
\cortext[cor1]{Corresponding author}
\cortext[cor2]{Ciudad Universitaria. Building 404, Room 343. Phone: (57)(1)3165000 Ext. 13051.}
\date{\today}

\begin{abstract}
In the context of a semiclassical approach where vectorial gauge
fields can be considered as classical fields, we obtain exact
static solutions of the $SU(N)$ Yang-Mills equations in a $(n+1)$
dimensional curved space-time, for the cases $n = 1, 2, 3$. As an
application of the results obtained for the case $n=3$, we
consider the solutions for the anti-de Sitter and Schwarzschild
metrics. We show that these solutions have a confining behavior
and can be considered as a first step in the study of the
corrections of the spectra of quarkonia in a curved background.
Since the solutions that we find in this work are valid also for
the group $U(1)$, the case $n=2$ is a description of the $(2+1)$
electrodynamics in presence of a point charge. For this case, the
solution has a confining behavior and can be considered as an
application of the planar electrodynamics in a curved space-time.
Finally we find that the solution for the case $n=1$ is invariant
under a parity transformation and has the form of
a linear confining solution.
\end{abstract}
\end{frontmatter}

\section{Introduction}

The quark confinement problem
\cite{conf11,conf12,conf13,conf14,conf15} cannot be solved by the
perturbation theory of quantum chromodynamics (QCD) so long as the
confinement is a nonperturbative phenomenon \cite{conf2}. In
essence the question is about how to confine the colour charges at
distances of order of characteristic hadronic sizes \cite{ymes1}.
The semiclassical approach permits to acquire non-perturbative
information about QCD starting from the solutions of classical
partial differential equations of $SU(3)$ Yang-Mills theory
\cite{ymes21,ymes22}. The searching of solutions of classical
Yang-Mills equations in presence of static external sources has
been motivated by understanding the role of non-abelian gauge
fields of quark confinement problem. One of the first works about
this subject showed that if an external source is distributed over
a thin spherical shell the Coulomb solution is unstable in a
specific regime \cite{ymes1}.

A wide range of solutions of $(3+1)$-dimensional Yang-Mills
equations in presence of localized and extended external sources
can be found in the literature \cite{ymes21}-\cite{ymes71}.
Likewise some exact retarded solutions to Yang-Mills equations
with sources composed of $N$ arbitrarily moving coloured point
particles were studied in \cite{kosy}. On the other hand some
specific solutions of the $(2+1)$-dimensional $SU(2)$ Yang-Mills
equations were obtained in \cite{d2ymes11,d2ymes12}. The discovery
of global regular solutions of $SU(2)$ Einstein-Yang-Mills
equations \cite{eymes1} originated a great interest about
spherical symmetric solutions \cite{eymes21}-\cite{eymes26}. For
instance, the solutions of the reduced $SU(2)$ Einstein-Yang-Mills
equations with spherical symmetry was presented in \cite{kunsle}.
Additionally some spherical solutions were considered in
\cite{winsta} for the $SU(2)$ Einstein-Yang-Mills theory with a
negative cosmological constant. Concurrently, it was given a
rigorous proof for the existence of infinitely many black hole
solutions to the $SU(3)$ Einstein-Yang-Mills equations
\cite{ruan}. Latterly, the study of quark confinement problem in
curved space-time has been a subject of interest
\cite{barros}.\par

A semiclassical approach motivated by the black hole physics
techniques was proposed by Goncharov \cite{yu0,yu00,yu4} to
describe the energy spectra of quarkonia by solving the Dirac
equation in presence of $SU(3)$ Yang-Mills fields representing
gluonic fields. In the context of this approach, explicit
calculations have shown that gluon concentration is huge at scales
of the order of $1$ fm \cite{yu1}. The solutions obtained there
can model the quark confinement satisfactorily, suggesting that
its mechanism might occur within the framework of QCD \cite{yu2}.
This implies that at large distances the gluons form a boson
condensate and, therefore, gluons can be described by the
classical $SU(3)$-gauge fields \cite{yu2}. For this reason the
dynamics of the strong interaction at long distances could be
described by equations of motion of the $SU(3)$ Yang-Mills theory
\cite{yu2}.
\par
In this work we follow the semiclassical approach proposed by
Goncharov, presented in detail in \cite{yu2}, and we obtain some
exact static solutions of the $(n+1)$-dimensional $SU(N)$
Yang-Mills equations in curved space-time, for the cases $n = 1,
2, 3$. As an application of the results obtained for the case
$n=3$, we consider the solutions for the anti-de Sitter and
Schwarzschild metrics. We show that these solutions have a
confining behavior and can be considered as a first step in the
study of the corrections of the spectra of quarkonia in a curved
background. Since the solutions that we find in this work are
valid also for the group $U(1)$, the case $n=2$ is a description
of the $2+1$ electrodynamics in presence of a point charge. For
this case, the solution has a confining behavior and can be
considered as an application of the planar electrodynamics in
curved space-time. i.e. electrodynamics in two spatial dimensions
(QED$_{2+1}$). For the case $n=1$, we find that the solution is
invariant under a parity transformation and have the form of a
linear confining solutions.
\par

The structure of this paper is as follows. In section 2 we first
present some preliminary aspects related to basic definitions and
notation. In section 3, we obtain the solutions for the
$(n+1)$-dimensional $SU(3)$ Yang-Mills equations in curved
space-time for the cases $n = 1, 2, 3$. Finally our conclusions
are summarized in section 4.


\section{Preliminaries}

We work on some spacetime manifold $M$ where in local coordinates the line
element looks as
\begin{equation} ds^2=g_{\mu \nu}dx^{\mu}dx^{\nu},
\end{equation}
and the components $g_{\mu \nu}$ take different values depending
on the choice of coordinates and dimensions. The \emph{Hodge start
operator} $*$ is defined by relation:
$\Lambda ^p(M)\rightarrow \Lambda ^{n-p}(M)$, where $\Lambda
^p(M)$ is the space of $\emph{p-form}$ over the manifold $M$ under consideration.
If $\{ dx^1,...,dx^n\}$ is the base for $\Lambda
^p(M)$ then
\begin{equation}
* (dx^{i_{1}}\wedge ...\wedge
dx^{i_{p}})=\frac{g^{(1/2)}}{(n-p)!}g^{i_1 l_1}...g^{i_p
l_p}\varepsilon_{l_1...l_pl_{p+1}...l_n}dx^{l_{p+1}}\wedge ...\wedge
dx^{l_{n}}.
\end{equation}
The \emph{exterior differential}, which is represented by $d$, is
defined as: $\Lambda ^p(M)\rightarrow \Lambda ^{p+1}(M)$. This
means that
\begin{equation}
d=\partial_\mu dx^{\mu}.
\end{equation}
If the connection $A$ in the gauge group $SU(N)$ is defined as
\begin{equation} \label{conA}
A=A_{\mu}dx^{\mu}=A_{\mu}^a \lambda _a dx^{\mu},
\end{equation}
where $A_{\mu}^a$ are the non-abelian fields associated to
$SU(N)$, $\lambda _a$ are the generators of $SU(N)$, with $a=1, 2,
..., N^2-1$, then the curvature $F$ can be defined using the
exterior differential as
\begin{equation}\label{defF}
 F=dA+gA\wedge A=F^a_{\mu \nu}\lambda _a dx^{\mu}\wedge dx^{\nu},
\end{equation}
where $F^a_{\mu \nu}$ represents the non-abelian stress tensor
associated to the gauge group $SU(N)$.

It is possible to write the $SU(N)$ Yang-Mills equations of the
non-abelian gauge field in presence of sources using the Hodge
star operator as follows
\begin{equation} \label{df}
 d*F=g(*F\wedge A-A\wedge *F)+gJ,
\end{equation}
where $F$ is the curvature (\ref{defF}), $A$ is the connection
(\ref{conA}), $g$ is the gauge coupling constant associated with
the gauge group $SU(N)$ and, for example, if considering the
QCD-lagrangian corresponding to that group then $J$ is a
nonabelian current given by
\begin{equation}
J=j^a_{\mu}\lambda _a*(dx^{\mu})=*j=*(j^a_{\mu}\lambda
_adx^{\mu})= \bar \Psi (I \otimes \gamma_\mu)\lambda^a\Psi \lambda
_a dx^\mu,
\end{equation}
where $\Psi$ are the Dirac fields, $\gamma^a$ are matrices which
represent the Clifford algebra and $I$ represents the identity.
For the case of a point particle at rest, this current density is
given by $J=*(j^a_{\mu}\lambda _a dx^{\mu})=\delta
(\vec{r})q^a\lambda _a*dt$, where $q^a$ are constants and then
$q^a\lambda _a=\Upsilon$ is a constant. In what follows we put
$J=\delta(\vec{r})q^a\lambda _a*dt$.
\par
In an analogous way as it is performed into the functional
quantization procedure of Yang-Mills theories, we fix the gauge
through the condition
\begin{equation}
\label{fijar} div(A)=\frac{1}{\sqrt{g}}\partial
_{\mu}(\sqrt{g}g^{\mu \nu}A_{\nu})=0.
\end{equation}


\section{Solutions in a $(n+1)$-dimensional curved space-time}

The $SU(N)$ Yang-Mills equations are a non-linear system of
coupled partial differential equations. In this section we will
present some static and exact solutions of the $SU(N)$ Yang-Mills
equations in a $(n+1)$ dimensional curved space-time, for the
cases $n= 1, 2, 3$.



\subsection{Case $n=3$}\label{subsec3}
The metric for a curved static space-time and spherical symmetry
is specified by
\begin{equation}
ds^2=g_{\mu \nu}dx^{\mu} dx^{\nu}=\alpha^2(r)dt^2-
\beta^2(r)dr^2-r^2(d \theta ^2+sin^2\theta d\varphi ^2).
\end{equation}
Due to we are interesting in the description of problems which
involve spherical symmetry in which the charge point is localized
in $r=0$, then we assume that the connection (\ref{conA}) has the
following functional dependence
\begin{equation}\label{ansatz}
A=A_{\mu}(r)dx^{\mu}=A_{t}(r)dt+A_{r}(r)dr+A_{\theta}(r)d\theta+
A_{\varphi}(r)d\varphi.
\end{equation}
For this case the gauge condition (\ref{fijar}) allow to the
following equation
\begin{equation}
\partial_r\left(\frac{r^2 \alpha ( r) A_r(r)}{\beta(r)}\right)+
\alpha ( r) \beta (r) A_{\theta}(r)cot\theta= 0,
\end{equation}
that can be satisfied if $A_{\theta}(r)=0$ and $A_r(r) = C
\frac{\beta(r)}{r^2 \alpha(r)}$. We can set $C=0$ in the $A_{r}$
solution because this fact does not affect the form of $A_{t}$ and
$A_{\varphi}$ solutions. So we assume the following anzats for the
form of these solutions \cite{San}
\begin{eqnarray}
A_{t}&=f(r)\Gamma,\label{ansatz1}\\
 A_{\varphi}&=g(r)\Delta,\label{ansatz2}
\end{eqnarray}
where $\Gamma$ and
$\Delta$ are linear combinations of the group generators. As the
exterior differential in spherical coordinates is
\begin{equation}
d=\partial_tdt+\partial_rdr+\partial_{\theta}d\theta+
\partial_{\varphi}d\varphi,
\end{equation}
then the curvature is given by
\begin{equation}\label{curcase1}
F=dA+gA \wedge A=-\partial _{r} f(r) \Gamma dt \wedge dr +
\partial _{r} g(r) \Delta dr \wedge d\varphi+g f(r) g(r)
[\Gamma ,\Delta ]dt \wedge d\varphi.
\end{equation}
For this case the $SU(N)$ Yang-Mills are written by the equations
(\ref{df}), but with the connection given by (\ref{ansatz}) and
the curvature by (\ref{curcase1}). Applying the Hodge start
operator to $F$ with inserting the result into (\ref{df}) and
using the following relations
 \begin{eqnarray*}
 * (dt\wedge dr)&=&-\frac{r^2  sin \theta}{\alpha(r) \beta(r)}
 d\theta \wedge d \varphi, \\
 * (dt\wedge d\theta )&=&\frac{\beta(r) sin \theta}{\alpha(r)} dr \wedge d \varphi,\\
 * (dt\wedge d\varphi)&=&\frac{-\beta(r)}{\alpha (r) sin\theta }dr \wedge d \theta, \\
 * (dr\wedge d\varphi )&=&\frac{-\alpha (r)}{\beta (r) sin\theta } dt \wedge d\theta ,\\
 * (dr\wedge d\theta )&=& \frac{\alpha (r) sin \theta}{\beta(r)} dt \wedge d \varphi ,\\
 * (d\theta \wedge d\varphi )&=&\frac{\alpha (r) \beta(r)}{r^2sin\theta } dt \wedge dr,
\end{eqnarray*}
we can obtain, for $r\not=0$, that the $SU(N)$ Yang-Mills
equations are reduced to the following coupled equation system
\begin{eqnarray}
\partial _{r}  \left( \frac{\alpha (r)\partial_r g(r)}{\beta (r) } \right) \Delta &=&
-\beta (r)\frac{g^2}{\alpha(r)} f(r)^2 g(r)[\Gamma ,[\Gamma ,\Delta ]],\label{23a}\\
\partial _{r}\left(\frac{r^2\partial _{r} f(r)}{\alpha(r) \beta(r)} \sin ^2
\theta\right) \Gamma &=&\frac{g^2 f(r)g(r)^2\beta(r)}{\alpha(r)}[\Delta
,[\Gamma ,\Delta ]]\label{23b}+g\frac{r^2\beta (r) \delta
(\vec{r})}{\alpha (r) }\Upsilon \sin ^2
\theta.
\end{eqnarray}
A non-trivial solution from this equation can be obtained if the
coupled equations (\ref{23a}) and (\ref{23b}) satisfy the abelian
condition $[\Delta ,[\Gamma,\Delta ]]=0$, then we obtain the two
following independent equations
\begin{eqnarray}
\partial _{r}  \left( \frac{\alpha (r)\partial_r g(r)}{\beta (r) } \right)&=&0,\\
\partial _{r}\left(\frac{r^2\partial _{r} f(r)}{\alpha(r) \beta(r)}\right)&=&g\frac{r^2\beta (0) \delta
(\vec{r})}{\alpha (0) }\Upsilon,
\end{eqnarray}
that have the following solutions
\begin{eqnarray}
g(r)&=&b_1 \int \frac{\beta(r)}{\alpha(r)}dr+B_1, \label{25}\\
f(r)&=&a_1 \int \frac{\alpha(r)\beta(r)}{r^2}dr+A_1. \label{24}
\end{eqnarray}
We note that the abelian condition $[\Gamma ,\Delta ]= 0$ is
satisfied in a non-trivial way if and only if one of the following
conditions hold:
\par 1) If within the combinations of $\Gamma$ and $\Delta$, in terms
of the generators of the group, it is present only the matrix that
constitutes the Cartan subalgebra of the $SU(N)$-Lie algebra i.e.
a maximal abelian subalgebra. This means that the commutator of any
two matrices of Cartan subalgebra is equal to zero.
\par 2) If $\Gamma=k \Delta$ and $k$ is a constant.


\subsubsection*{C.1. Case of flat space-time}

We can obtain from (\ref{25}) and (\ref{24}) that the solutions,
for the flat space-time case $\alpha=\beta=1$, are
\begin{equation} \label{cs3d}
g(r)=b_1 r +B_1, \,\,\,\,\ f(r)= -a_1/r +A_1.
\end{equation}
These solutions are confining solutions in a semiclassical
approach where the gluons can be considered as classical fields
\cite{yu0,yu00,yu4,yu1,yu2,yu3,yu51,yu52,yu53}. We have obtained the
confining solutions (\ref{cs3d}) using the ansatz given by
(\ref{ansatz1}) and (\ref{ansatz2}). We note that these solutions
were obtained in \cite{yu00}, but using the ansatz
\begin{eqnarray}
A=r^{\mu_a}\alpha^a\lambda_a dt+A_rdr+r^{\rho_a}\gamma^a\lambda_a d
\theta+r^{\nu_a}\beta^a\lambda_a d\varphi,
\end{eqnarray}
where $\mu_a$, $\nu_a$, $\rho_a$, $\alpha^a$, $\gamma^a$ and
$\beta^a$ are arbitrary real constants. The confining potentials
between quarks are usually modeled as $V(r)=A/r+Br$, however it
has been shown that $A_tdt=(A/r+Br)dt$ is not a solution of the
Yang-Mills equations in the presence of a point charge \cite{yu3}.
The solutions (\ref{cs3d}), that we have found here, were applied
to describe the energy spectra of quarkonia (charmonium and
bottomonium) in \cite{yu4}, to predict the electric form factor,
the magnetic moment and the root-mean-square radius of mesons in
\cite{yu51,yu52,yu53}, and to study the chiral symmetry breaking in QCD
\cite{yu6}.\par

It was shown that the confining solutions (\ref{cs3d}) satisfy the
so-called Wilson confinement criterion \cite{yu2}. This criterion
is in essence the assertion that the so-called Wilson loop $W(c)$
should be subjected to the area law for the confining gluonic
field configuration. As a consequence, the latter law is
equivalent to the fact that energy $E(R)$ of the mentioned
configuration (gluon condensate) is linearly increasing with $R$,
a characteristic size of some volume $V$ containing the condensate
\cite{yu2}. The evaluation of $E(R)$ was carried out in \cite{yu2}
using the $T_{00}$ - component of the energy-momentum tensor for a
$SU(3)$-Yang-Mills field. However, formally $E(R)$ diverges for
everything $R$. Calculations carried out in \cite{yu2} have
considered the integral about the angle $\theta$ inside the limits
$(\theta_0$,$\pi-\theta_0)$, with the purpose of avoiding the
divergence. Next we will consider a classical estimate of Wilson
loop in the real-time formalism ($\tau\rightarrow it$) directly
for the gluon condensate, i. e.
\begin{eqnarray}
W(R,T)\equiv  < e^{ig\oint A_{\mu}dx^{\mu}}>=e^{-T V(R)},
\end{eqnarray}
and taking into account that the expectation value coincides with
the evaluation of the integral in this approach. We obtain as
result that
\begin{eqnarray}
\oint A_{\mu} dx^{\mu}=\int_0^{2\pi }(b_1 R +B_1)Tr\Delta
d\varphi=2\pi(b_1 R +B_1)Tr\Delta,
\end{eqnarray}
where $Tr$ is the trace of the matrix. We observe from this result that the confinement potential
obtained has the form $V(R)=\sigma R+C$. We note that this
solution has the form of the confining Cornell potential
\cite{cornell} which has been used to describe experimental
features of QCD. Lattice QCD simulations carried out have found
the same kind of potential from Wilson loop \cite{Ynd,Gre,wal}.


\subsubsection*{C.2. Case of anti-de Sitter metric}

As a particular application of the $(3 + 1)$ case solution, we
will consider the anti-de Sitter metric given by
\begin{equation}
ds^2=(1-\Lambda r^2/3)dt^2-(1-\Lambda r^2/3)^{-1}dr^2-r^2 (d \theta
^2+\sin^2\theta d\varphi ^2).
\end{equation}
As $\alpha(r)$ is the inverse of $\beta(r)$, then the Coulomb
solution has not deformations respect to the flat space-time case,
then the solution of the function $f(r)$ is given by
$f(r)=a_1/r+A_1$. On the other hand, the linear solution $g(r)$
changes respect to the flat case. We get the solution explicitly
in the two following situations $\Lambda>0 $ and $\Lambda <0 $, so
\begin{equation} \label{smcs}
g(r) = \left\{
\begin{array}{cl}
b_1 \tanh ^{-1}\left(\frac{r \Lambda^{1/2}}{3^{1/2}}\right)
+B_1,&\mbox{if } \Lambda >0,\\
b_1 \tan ^{-1}\left(\frac{r (-\Lambda)^{1/2}}{3^{1/2}}\right)
+B_1,&\mbox{if } \Lambda <0,\\
\end{array}\right.
\end{equation}

The function $g(r)$, for the limit cases $|\Lambda|<<1$ and $r<<1$,
has the form $g(r)\simeq b_1r+B_1$, recovering the flat space-time
case behavior.
\par


\subsubsection*{C.3. Case of Schwarzschild metric}

Another application for the $(3 + 1)$ curved space-time solution is
the Schwarzschild metric given by
\begin{equation}\label{shr}
ds^2=(1-2M/r)dt^2-(1-2M/r)^{-1}dr^2-r^2(d \theta ^2+ sin^2\theta
d\varphi ^2).
\end{equation}
As in the last application, the Coulomb solution has not
deformations respect to the flat space-time case, but the linear
solution has. The function $g(r)$ for this case is
\begin{equation} \label{scmcs}
g(r)=b_1 (r + 2 M \ln|r-2 M|)+B_1.
\end{equation}
For the limit $r>>M$, this solution has the form $g(r)\simeq
b_1r+B_1$, recovering the flat space-time case behavior.

We observe that the solutions (\ref{smcs}) and (\ref{scmcs}) are
confining solutions. From the analysis of the spectra of quarkonia
in a spherical symmetry background, we can observe that the linear
solution changes as a consequence of the curvature. This fact
could generate a change in the mass spectrum of mesons. This
change additionally could be generated by the the effects of the
curvature in the Dirac equation. By this reason, the solutions
(\ref{25}) and (\ref{24}) can represent a first step in the study
of hadronic spectrum in a space-time curved.


\subsection{Case $n=2$}
We consider now a curved space-time in $(2 + 1)$ dimensions
defined by the metrics
\begin{equation}
ds^2=g_{\mu \nu}dx^{\mu} dx^{\nu}=\alpha^2(r)dt^2-\beta^2(r)
dr^2-r^2d\theta^2.
\end{equation}
For this case we assume that the connection (\ref{conA}) has the
following form
\begin{equation} \label{asemA}
A = A_{\mu}(r)dx^{\mu} = A_t (r)dt+ A_r
(r)dr+A_{\theta}(r,\theta)d\theta.
\end{equation}
Using the gauge condition (\ref{fijar}), we can obtain the
following equation
\begin{equation}\label{fijotra}
\partial_r\left(\frac{r \alpha ( r) A_r(r)}{\beta(r)}\right)+
\partial_{\theta}\left(\frac{\alpha ( r) \beta (r) A_{\theta}
(r,\theta)}{r}\right)= 0.
\end{equation}
In the last equation, the solutions of the form $A_{\theta}(r,
\theta) = H(r)\theta + G(r)$ can be discarded because they must be
periodic in $\theta$, i. e. $A_{\theta}$ does not depend on
$\theta$. Thus, the equation (\ref{fijotra}) can be written as
\begin{equation}
\partial_r\left(\frac{r \alpha ( r) A_r(r)}{\beta(r)}\right)=0.
\end{equation}
The solution of this equation has the form $A_r = C
\frac{\beta(r)}{r \alpha(r)}$. In these coordinates, the exterior
differential is written as
\begin{equation}\label{d22}
d = \partial_t dt + \partial_rdr+\partial_{\theta}d\theta.
\end{equation}
If now we set $C = 0$ and we substitute the functions $A_t =
f(r)\Gamma$ and $A_\theta = g(r)\Delta$ in (\ref{asemA}), where
$\Gamma$ and $\Delta$ are linear combinations of the group
generators, and additionally we consider the following relations
 \begin{eqnarray*}
 * (dt\wedge dr)&=&-\frac{r}{\alpha (r) \beta (r)} d\theta, \\
 * (dt\wedge d\theta )&=&\frac{\alpha (r) \beta (r)}{r}dr,\\
 * (dr\wedge d\theta )&=&\frac{\alpha (r)}{r \beta (r)}dt,
\end{eqnarray*}
then the curvature is given by
\begin{equation}\label{c22}
F=dA+gA \wedge A=-\partial _{r} f(r) \Gamma dt \wedge dr +
\partial _{r} g(r) \Delta dr \wedge d\varphi+
g f(r) g(r)[\Gamma ,\Delta ]dt \wedge d\theta,
\end{equation}
and the Hodge star operator applied over the curvature can be
written as
\begin{equation}
* F=\frac{r\partial _{r} f(r) \Gamma}{\alpha (r) \beta (r)} d\theta+
\frac{\alpha (r)}{r \beta (r)}\frac{\partial _{r} g(r)}{r} \Delta dt
+ g \frac{\alpha (r) \beta (r) f(r)g(r)}{r}[\Gamma ,\Delta ]dr.
\end{equation}
The Yang-Mills equation for this case can be obtained from the
expression (\ref{df}) and has the form
\begin{eqnarray}  \label{eqtr}
&&\partial_r\left(r \frac{\partial_rf(r)}{\alpha(r)\beta(r)}\Gamma
\right) dr \wedge d\theta - \partial_r\left(\frac{ \alpha(r)
\partial_rg(r)}{r \beta(r)}\Delta \right) dt \wedge dr \,\, = \,\,
g\delta (\vec{r}) \frac{r \beta(r)}{\alpha(r)} \Upsilon dr
\wedge d\theta \nonumber \\
&+&g^2\left( \frac{f(r)g(r)^2\alpha(r)\beta(r)}{r}[[\Gamma,\Delta],
\Delta] dr \wedge d\theta - \frac{f(r)^2g(r)\alpha(r)\beta(r)}{r}
[[\Gamma,\Delta],\Gamma] dt \wedge dr
 \right).\nonumber\\
\end{eqnarray}
If the abelian condition is satisfied in the last equation, i. e.
the commutator $[\Gamma ,\Delta ]= 0$, then the solutions will be
strongly restricted by this condition. Thus, we can obtain the two
following independent equations
\begin{eqnarray}
\partial_r \left(\frac{\alpha(r)}{r \beta (r)}\partial_r g(r)
\right) &=& 0,\\
\partial _{r}\left( \frac{r}{\alpha (r) \beta (r)}\partial _{r}
f(r) \right)&=& g\delta (\vec{r}) \frac{r \beta(0)}{\alpha(0)}
\Upsilon,
\end{eqnarray}
which solutions are given by
\begin{eqnarray}
g(r)&=&k_2 \int \frac{r \beta (r)}{\alpha (r)}dr+d_2,\\
f(r)&=&k_1 \int \frac{\alpha (r) \beta (r)}{r}dr+d_1.
\end{eqnarray}
For $\alpha=\beta=1$, these solutions can be written as
\begin{eqnarray}
g(r) = d_2+k_2r^2,\\
f(r) = d_1+k_1 \log r.
\end{eqnarray}
We observe that the function $f(r)$ has the form of the well known
confining solution for the two dimensional problem \cite{arfken}.
Similarly, the function $g(r)$ has a form of a confining function.
It is possible to eliminate the constant $d_1$ by means of a gauge
transformation, so the potential is given by
\begin{equation}
A_{\mu}dx^{\mu}=k_1 \Gamma \log r dt+(d_2+k_2r^2)\Delta d\varphi,
\end{equation}
and the $1$-form vectorial field is
\begin{equation}\label{signi}
A_{\mu}=\left\{k_1 \Gamma \log r ,0,(\frac{d_2}{r}+k_2 r)\Delta\right
\}.
\end{equation}
In the last expression, the term $k_1 \Gamma \log r$ is a
potential equivalent to the Coulomb potential in $(2+1)$
dimensions and additionally it corresponds to the potential which
is obtained for an infinite line of charge in $(3+1)$ dimensions.
Additionally, the term $k_2 r$ in the $z$ axis is equivalent to a
constant magnetic field of magnitude $B_1=2 k_2$ in $(3+1)$
dimensions. We can see that if we perform a transformation to
cartesian coordinates over this last term, we obtain its usual
representation
\begin{equation}
k_2 r \hat{e}_{\varphi}=\frac{B_1}{2} r (-\sin \varphi\hat{e}_x+\cos
\varphi\hat{e}_y)=-\frac{B_1}{2} y \hat{e}_x+\frac{B_1}{2}x\hat{e}_y.
\end{equation}
Finally, the term $\frac{d_2}{r}$ can be interpreted as the
potential produced by an infinite solenoid of radius $R$ in
$(3+1)$ dimensions, with $R\rightarrow 0$. This last term can be
written in its usual representation by means of a transformation
to cartesian coordinates
\begin{equation}
\frac{d_2}{r} \hat{e}_{\varphi}=\frac{d_2}{r} (-\sin \varphi\hat{e}_x+
\cos \varphi\hat{e}_y)=-\frac{d_2}{r^2}y \hat{e}_x+\frac{d_2}{r^2}x\hat{e}_y.
\end{equation}
This result is known as the Aharonov-Bohm (AB) potential, where
the magnetic field within the solenoid is $B_2=d_2$.

Let's consider now the integral that appears in the Wilson loop
\begin{equation}
\oint A_{\mu}dx^{\mu}.
\end{equation}
This integral can be performed in a circular path of radius $R$
which is centered in the origin of the coordinates frame. Then we
obtain that
\begin{equation}
\oint A_{\mu}dx^{\mu}=\int_0^{2\pi }(d_2+k_2R^2)Tr\Delta d\varphi =
2 \pi(d_2+k_2R^2) Tr\Delta.
\end{equation}
This result implies that
for this approach the confinement potential in low dimensionality
is given by $V(R)= \sigma R^2+B$.

The result obtained for this case correspond to a $U(1)$ abelian
solution which has a confining behavior and can be considered as
an application of the planar electrodynamics, i.e. electrodynamics
in two spatial dimensions (QED$_{2+1}$). QED$_{2+1}$ has been
worked in many emblem quantum systems such as the quantum Hall
effect, the theory of anyons and the relativistic quantum Hall
effect \cite{Pran,Wilc,AMJ}. During the last years the physics of
graphene has attracted considerable interest, both theoretically
and experimentally \cite{Geim,Neto}. Due to the low energy
excitations of graphene can be described by a massless Dirac
equation in two spacial dimensions, the curved graphene has been
modeled by coupling the Dirac equation to the corresponding curved
space \cite{Voz1}. In the approach presented in \cite{Voz2}, gauge
fields has been considered as external fields into the Dirac
equation and it was possible to model some topological defects in
the graphene. In connection with the results that we present in
this work, the solutions that we obtain for the case $n=2$ can be
interpreted as a gauge field which is affected by the curvature of
graphene.

\subsection{Case $n=1$}

For this case we start from the metric given by
\begin{equation}
ds^2=g_{\mu \nu}dx^{\mu}
dx^{\nu}=\alpha^2(x)dt^2-\beta^2(x)dx^2.
\end{equation}
We suppose that the connection has the following functional
dependence
\begin{equation}
A = A_{\mu}(x) dx^{\mu} = A_t (x) dt+ A_x(x) dx=\lambda_a f^a(x) dt
+ A_x(x) dx,
\end{equation}
where $\lambda_a$ are the group generators. Each of these
generators has associated a function $f^a(x)$. The gauge condition
(\ref{fijar}) implies that
\begin{equation}
\partial_x\left(\frac{\alpha (x) A_x(x)}{\beta(x)}\right)= 0,
\end{equation}
and then $A_x = C \beta(x)/\alpha(x)$. On this coordinates, the
exterior differential is written as
\begin{equation}\label{d11}
d = \partial_t dt + \partial_xdx.
\end{equation}
Fixing $C=0$ and using the fact that
$*(dt\wedge dx)=-\frac{1}{\alpha (x) \beta (x)} $, it is possible to
write
\begin{equation}
*F=\frac{\lambda_a\partial_xf^a(x)}{\alpha(x)\beta(x)}.
\end{equation}
For this case we have that $*(dt)=\frac{\beta(x)}{\alpha(x)}dx$
and the expression (\ref{df}) leads us to
\begin{eqnarray}
\partial_x\left(\frac{\lambda_a\partial_xf^a(x)}{\alpha(x)\beta(x)}
\right)dx=g\left(\frac{\partial_xf^a(x)}{\alpha(x)\beta(x)}f^b(x)
[\lambda_a,\lambda_b]\right)dt+g\delta(x) \frac{\beta(x)}{\alpha(x)}
q^a\lambda_adx.
\end{eqnarray}
A non-trivial solution from this equation can be obtained if the
condition $[\lambda_a,\lambda_b]=0$ is satisfied, or alternatively
$\lambda_a\partial_xf^a(x)=\partial_xf(x)\Gamma$, where $\Gamma$ represents
a linear combination of the group generators. Thus, the last
equation can be written as
\begin{eqnarray}
\partial_x\left(\frac{\lambda_a\partial_xf^a(x)}{\alpha(x)\beta(x)}
\right)=-g\delta(x) \frac{\beta(0)}{\alpha(0)}q^a\lambda_a.
\end{eqnarray}
We solve this equation and obtain
\begin{eqnarray}
f^a(x)=k^a\int \alpha (x) \beta (x) dx +d^a,
\end{eqnarray}
where $k^a$ and $d^a$ are constants. For $\alpha=\beta=1$ we can
write $f(x) = d + k|x|$, i. e. we have obtained the solution of
the flat space-time case. This last solution is invariant under a
parity transformation and has the form of a linear confining
solution. The form of this solution corresponds to the potential
that results due to an infinite sheet of charge in $(3+1)$
dimensions.



\section{Conclusions}
We have obtained some exact static solutions for the $SU(N)$
Yang-Mills equations in a curved space-time of $(n+1)$ dimensions
for the cases $n = 1, 2, 3$. For the $(1+1)$ case, we have found
that the solution for the temporary part can be written as
$A_t=f(r)\Gamma$. This solution is the most general static
solution in a space-time in presence of a point charge. To be able
to find analytic solutions in the $(2+1)$ case, it was necessary
to demand that the abelian condition given by $[\Delta,\Gamma]=0$
were satisfied into the Yang-Mills equation for this case. For the
cases $(1+1)$ and $(3+1)$, the abelian condition was naturally
satisfied. We have presented in detail the solution for $(3+1)$
curved space-time case and we have applied this solution to the
anti-de Sitter and Schwarzschild cases. In both cases, the Coulomb
solutions have not deformations respect to the flat space-time
case, while in the linear solution exists deformation. We have
assumed that these solutions can be considered as a first step in
the study of the corrections on the spectra of quarkonia in a
curved background. Although the energy diverges for the solutions
that we have found, this fact is not a problem since this energy
behavior was already known and it is not an impediment to make
physical predictions in agreement with experimental data
\cite{yu4,yu3,yu51,yu52,yu53,yu6}. Since the solutions that we have found
here are valid also for the group $U(1)$, the case $n=2$ is a
description of the $(2+1)$ electrodynamics in presence of a point
charge. Finally, we have found that the solution for the case
$n=1$ is invariant under a parity transformation and has the form
of a linear confining solution. As a perspective of this work, it
would be interesting to understand the role of the confining
solutions in a model of relativistic quark confinement in low
dimensionality and the spectra of quarkonia in a curved
background.

\section*{Acknowledgments} We are indebted to Maurizio de Sanctis
and Rafael Hurtado for the suggestions about the elaboration of
the present work. C. J. Quimbay would like to thank Vicerrectoria
de Investigaciones of Universidad Nacional de Colombia by the
financial support received through the research grant ``Teor\'ia
de Campos Cu\'anticos aplicada a sistemas de la F\'isica de
Part\'iculas, de la F\'isica de la Materia Condensada y a la
descripci\'on de propiedades del grafeno''.

\end{document}